\documentclass[%
 reprint,
 amsmath,amssymb,
 aps,
]{revtex4-1}
\newcommand{\sx}{\hat{\sigma}_x}
\newcommand{\sy}{\hat{\sigma}_y}
\newcommand{\sz}{\hat{\sigma}_z}

\usepackage{graphicx}
\usepackage[braket, qm]{qcircuit}
\usepackage[utf8]{inputenc}

\begin{document}

\title{Measuring the Mermin-Peres magic square using an online quantum computer}
\author{A Dikme}
\author{N Reichel}
\author{A Laghaout}
\author{G Bj\"{o}rk}
\email{gbjork@kth.se}
\address{Department of Applied Physics, Royal Institute of Technology (KTH), \\
Albanova University Center, 106 91 Stockholm,\\ Sweden.}

\begin{abstract}
We have implemented the six series of three commuting measurement of the Mermin-Peres magic square on an online, five qubit, quantum computer. The magic square tests if the measurements of the system can be described by physical realism (in the EPR sense) and simultaneously are non-contextual. We find that our measurement results violate any realistic and non-contextual  model by almost 28 standard deviations. We also find that although the quantum computer we used for the measurements leaves much to be desired in producing accurate and reproducible results, the simplicity, the ease of re-running the measurement programs, and the user friendliness compensates for this fact.
\end{abstract}

\maketitle

\section{Introduction}

The much cited EPR-paper \cite{ EPR} was borne out of a conviction on the authors' part that a reasonable physical theory must be realistic. In the words of EPR, realism means that: "If, without in any way disturbing a system we can predict with certainty (i.e., with probability equal to unity) the value of a physical quantity, then there exists an element of physical reality corresponding to this physical quantity''. As reasonable as realism seems, it is well known by now that if additional conditions are imposed, such as non-locality \cite{Bell,Hensen}, non-signalling \cite{Scarani}, or non-contextuality \cite{Kochen,Peres,Mermin}, quantum mechanics readily violates models based on these conditions. In this paper we specifically use an experimental test that shows that quantum theory is at odds with theories that assume realism and non-contextuality.

Typically such tests are difficult to perform because either the tests require particular (typically entangled) initial states, and in addition they require rather elaborate measurements. Thus, it is relatively recently experimental groups worldwide have had the resources and know-how to perform these rather delicate measurements with a precision and accuracy required to verify that quantum mechanics is at odds with the combination of realism and non-contextuality \cite{Huang,Bartosik,Kirchmair,Liu,Moussa,Amselem,Hasegawa,Lapkiewicz,Zu,Dogra,Canas,Crespi,Xiao,Zhang}. Just as with Bell- \cite{Bell} and Leggett-Garg \cite{Leggett} types of tests, experimental imperfections typically ``steer'' the results in the direction of possible realistic, local, and non-contextual explanation models. To facilitate experimental tests, a number of state-independent inequalities testing contextuality have been proposed \cite{Cabello2008}. The test we have done, the Mermin-Peres magic square, is such a test. It was performed for the first time in 2009, using an elaborate ion-trap setup \cite{Kirchmair}. The experiment showed a convincing violation of any theory that is both realistic and non-contextual. To the best of our knowledge this specific experimental test has not been repeated, presumably because it is experimentally very challenging, although it is state-independent. That is, no elaborate input state is needed, the test ideally yields the same violation regardless of two-qubit input state.

However, recently the quest for user-friendly quantum computers have opened up an attractive possibility for doing rather complicated measurements using the qubits in a quantum computer as the experimental quantum system, at least for measurement problems that are naturally mapped onto qubits. There is still a very limited number of operating quantum computers worldwide, but luckily, some of them are available openly and free of charge via the internet. They typically operate in ``batch mode''. That is, one writes a program, uploads it via the internet to a queue, waits for the program to run on the remote quantum computer (typically it is run more than once due to imperfect hardware), and finally one receives the result file with the (measurement) outcomes of the runs. This enables students, researchers on a tight budget, and theoreticians to test quantum propositions experimentally without having to make the (often expensive and very time consuming) effort of building their own experimental set-up. Examples are given in \cite{Alsina,Devitt,Garcia,Smart}.

In this way we have tested the Mermin-Peres magic square \cite{Mermin}, an experiment that requires nine different measurements on a two-qubit system. What makes this experiment challenging is that one needs to make sequences of three different, sequential measurements on the very same system. This means that the measurements must not be destructive, but be of the quantum non-demolitional (QND) kind \cite{Braginsky}. In general this requires the system to become entangled with a probe (in our case an ancilla qubit) that can subsequently be measured demolitionally (in general causing the state of the measured system to change). Moreover, the observables belonging to the magic square have degenerate eigenvalues. As a consequence, the states that the five two-qubit measurements collapse the state to are in general entangled. Such states are fragile in themselves, but to complete all the measurements needed for the magic square one will have to make one or two subsequent QND-measurements of the entangled system. Anyone familiar with experimental quantum physics will testify that this is (still) a rather challenging task. However, by using the steadily improved qubit technology offered through quantum computers, such an experiment is now at the hand of anyone having access to an internet connected quantum computer, albeit with non-negligible measurement noise added.

\section{The Mermin-Peres magic square}

The Mermin-Peres magic square was suggested by Mermin \cite{Mermin}, based on earlier work by Peres \cite{Peres}, as a scheme to test the simultaneous hypotheses that quantum mechanics could be described by a realistic and non-contextual model. The square consists of a 3 x 3 matrix, where each of the nine entries is an observable operating on two qubits, (or originally, on two spin 1/2 systems). The observables consist of tensor products of the Pauli operators $\sx$, $\sy$,  $\sz$, and the identity operator, see table \ref{Table 1}. The units are chosen so that the outcome of any of the nine measurements is either +1 or -1, and the square is ingeniously arranged such that the product of the three measurement outcomes of any of the three rows, or any of the three columns, is deterministic irrespective of the initial state of the two qubits, according to quantum mechanics.
\begin{table}
\caption{\label{Table 1} The Mermin-Peres magic square. The numbers to the right and under the square is the deterministic, quantum mechanically predicted products of the three measurement outcomes for each row and each column.}
\begin{tabular}{||c|c|c||r}
  \hline\hline
  $\sx \otimes 1$ & $1 \otimes \sx$ & $\sx \otimes \sx$ & 1 \\
    \hline
  $1 \otimes \sy$ & $\sy \otimes 1$ & $\sy \otimes \sy$ & 1 \\
  \hline
  $\sx \otimes \sy$ & $\sy \otimes \sx$ & $\sz \otimes \sz$ & 1 \\
  \hline\hline
  1 & 1 & -1 &  \\
\end{tabular}
\end{table}
The products of the rows' and the columns' observables can be checked using the Pauli operator commutation relations. It can also be checked that the three measurements in each row and in each column commute.

If one assumes that the measurements of the square can be described by a realistic and non-contextual theory, then one should be able to model any experimental outcome by simply assigning the numbers +1 or -1 to each and every of the nine observables. There are $2^9 = 512$ different ways of doing this, and the measurement results should, if the measurement results all were in fact +1 or -1, be possible to model by a suitable statistical mixture of such outcome matrices. By simple trial and error one can quickly establish that it is not possible to fill in the numbers +1 and -1 in the 3 x 3 square such that all the six products come out as predicted by quantum mechanics. The reason is that in order to get the product 1, one needs an even number of -1 entries in the rows or columns. Thus to get the three rows all to have the product 1, the total number of -1s must be an even number. The same holds for the two leftmost columns, but for the rightmost columns one needs an odd number of -1s. Thus, to satisfy all three column product results, the number of -1s in the square must be odd. One is led to a contradiction, so at least one of the underlying assumptions must be incorrect. The three assumptions leading to this contradiction were 1) that the measurements in the square could be explained by a realistic theory. That is, to each measurement one should be able to assign an outcome, and in the case the outcome is non-deterministic, the measurement results should be explainable by a statistical mixture of assigned outcomes. 2) That the measurements are non-contextual. Thus, for a given square, the assigned outcome in any square is independent of if it is measured as part of a row sequence or a column sequence. 3) That quantum mechanics accurately predicts the (deterministic) measurement outcome products (but not necessarily the outcomes of every measurement since for any given input state, some of them will be non-deterministic according to quantum theory).

Since we are only interested in the measurement products, and these can also only be +1 or -1, one needs only consider the $2^6 = 64$ possible product outcomes. In the following, we will write these as a six-component vector with the products of the measurements of the rows, top to bottom, as coefficients 1 to 3, respectively, and the products of the measurements of the columns, left to right, as coefficients 4 to 6, respectively. We will refer to such vectors as result vectors. Thus, the result vector of table \ref{Table 1} is $(1,1,1,1,1,-1)$. The 64 result vectors can be divided into two groups of 32. The ``realism'' group consists of vectors that are commensurable with realism and non-contextuality. They have to consist of entries that assume either an even or an odd number of -1s in the square. If we look at either the three rows or the three columns, the three products can be sorted as in table \ref{Table 2}.
\begin{table}
\caption{\label{Table 2} The possible $2^3$=8 measurement products of the three rows, or the three columns, sorted in two groups. The even group requires an even number of -1s in the square, and the odd group requires an odd number of -1s.}
\begin{tabular}{|c|c|}
  \hline
  Even & Odd \\
  \hline
  $(1,1,1)$ & $(-1,-1,-1)$ \\
  $(1,-1,-1)$ & $(-1,1,1)$ \\
  $(-1,1,-1)$ & $(1,-1,1)$ \\
  $(-1,-1,1)$ & $(1,1,-1)$ \\
  \hline
\end{tabular}
\end{table}

The group of ``realism'' result vectors must be composed of either any two entries from the even group or from any two entries from the odd group of table \ref{Table 2}, depending on whether the number of -1s in the square is even or odd. E.g., taking $(1,1,1)$ and $(1,-1,-1)$ from the even group, we can form the two result vectors $(1,1,1) \oplus (1,-1,-1)= (1,1,1,1,-1,-1)$ or $(1,-1,-1) \oplus (1,1,1)= (1,-1,-1,1,1,1)$. Both these result vectors are generated by realistic and non-contextual squares containing an even number of, and at least two -1s, see table \ref{Table 4} for examples.
\begin{table}
\caption{\label{Table 4} Two realistic magic squares containing two and six -1s, respectively. The squares generate the result vectors $(1,1,1,1,-1,-1)$ and $(1,-1,-1,1,1,1)$, respectively.}
\begin{tabular}{||c|c|c||r}
  \hline\hline
  1 & 1 & 1 & 1 \\
    \hline
  1 & -1 & -1 & 1 \\
  \hline
  1 & 1 & 1 & 1 \\
  \hline\hline
  1 & -1 & -1 &  \\
\end{tabular}
\quad \quad
\begin{tabular}{||c|c|c||r}
  \hline\hline
  -1 & -1 & 1 & 1 \\
    \hline
  -1 & -1 & -1 & -1 \\
  \hline
  1 & 1 & -1 & -1 \\
  \hline\hline
  1 & 1 & 1 &  \\
\end{tabular}
\end{table}

The total number of such ``realism'' vectors one can form is $4 \times 4 + 4 \times 4 = 32$. The vectors in this group will be denoted $\overline{r}_1$, $\overline{r}_2$, \ldots, $\overline{r}_{32}$ where the ordering is inconsequential. If, on the other hand, one combines one entry from the even group with one entry from the odd group, then one obtains a result vector that are at odds with realism and non-contextuality. One such vector is $(1,1,1) \oplus (1,1,-1) = (1,1,1,1,1,-1)$, that is, the outcome predicted by quantum mechanics for the magic square. We will call this vector $\overline{q}_1$, and the remainder of the 32 vectors in this group will be denoted $\overline{q}_2$, $\overline{q}_3$, \ldots, $\overline{q}_{32}$, where again, the ordering is unimportant for what follows.

\section{Measuring the magic square}
To perform the Mermin-Peres magic square measurements we have used the quantum computers ibmqx2 ``Yorktown'', ibmqx4 ``Tenerife'', and ibmqx16 ``Melbourne'', all publicly available through the IBM Q Experience platform on the internet \cite{IBM quantum experience}. Of the three, we got the best results (the results that were in the best agreement with the quantum mechanical predictions) on the ibmqx4 computer. All the computers we used fit into the category ``noisy, intermediate scale quantum'' (NISQ) computers. The former two are perhaps not even intermediate scale, but small scale.

The ibmqx4 has five qubits, interconnected in a butterfly pattern. Qubit 0, 1, and 2 are fully interconnected, and so are qubits 2, 3, and 4. Thus, only qubit 2 is connected to all the remaining four. The decoherence times $T_1$ and $T_2$ for the different qubits vary between 30.1 to 52.1 $\mu$s and 4.9 to 53.0 $\mu$s, respectively. The single and two-qubit gate-errors likewise show a significant spread. The one-qubit gates have errors between 0.69 to 3.37 per mille, while the two-qubit gate-errors lie in the range 2.12 to 7.95 percent. Finally there are qubit readout errors that, depending on the qubit, ranges from 3.4 to 34.8 (!) percent.

\begin{figure}[h]
\[
 \hspace{14mm}
    \Qcircuit @C=1.27em @R=0.2em @!R {
	 	\lstick{q_{0}: \ket{+1}} & \gate{H} & \ctrl{2} & \qw & \ctrl{2} & \qw & \qw & \qw& \qw\\
	 	\lstick{q_{1}: \ket{+1}} & \gate{H} & \qw & \ctrl{1} & \qw & \ctrl{1} & \qw & \qw& \qw\\
	 	\lstick{q_{2}: \ket{+1}} & \qw & \targ & \targ & \targ & \targ & \gate{M_z} & \qw& \qw\\
	 }
 \]
\vspace{2mm}
\[
   \hspace{14mm}
    \Qcircuit @C=.85em @R=.2em  {
	 	\lstick{q_{0}: \ket{+1}} & \gate{u_{2}(0,\frac{\pi}{2})} & \qw & \ctrl{2} & \qw & \ctrl{2} & \qw & \qw & \qw& \qw\\
	 	\lstick{q_{1}: \ket{+1}} & \gate{u_{2}(0,\frac{\pi}{2})} & \qw & \qw & \ctrl{1} & \qw & \ctrl{1} &\qw & \qw & \qw\\
	 	\lstick{q_{2}: \ket{+1}} & \qw & \qw & \targ & \targ & \targ & \targ & \gate{M_z} & \qw & \qw\\
	 }
 \]
\vspace{2mm}
\[
  \hspace{14mm}
    \Qcircuit @C=.1em @R=0.2em {
	 	\lstick{q_{0}: \ket{+1}} & \ctrl{2} & \qw & \gate{H} & \ctrl{2} & \qw & \gate{u_{2}(\frac{\pi}{2},\frac{3\pi}{2})} & \qw & \ctrl{2} & \qw & \qw & \qw & \qw \\
	 	\lstick{q_{1}: \ket{+1}} & \qw & \ctrl{1} & \gate{u_{2}(0,\frac{\pi}{2})} & \qw & \ctrl{1} &  \gate{u_{2}(\frac{3\pi}{2},\frac{\pi}{2})}  & \qw & \qw & \ctrl{1} & \qw & \qw \\
	 	\lstick{q_{2}: \ket{+1}} & \targ & \targ & \qw & \targ & \targ & \qw & \qw & \targ & \targ & \gate{M_z} & \qw & \qw \\
	 }
 \]
\vspace{2mm}
\[
  \hspace{14mm}
    \Qcircuit @C=.83em @R=0.3em @!R {
	 	\lstick{q_{0}: \ket{+1}} & \gate{H} & \ctrl{2} & \qw & \ctrl{2} & \qw & \qw & \qw & \qw& \qw\\
	 	\lstick{q_{1}: \ket{+1}} & \gate{u_{2}(0,\frac{\pi}{2})} & \qw & \ctrl{1} & \qw & \ctrl{1} & \qw & \qw & \qw& \qw\\
	 	\lstick{q_{2}: \ket{+1}} & \qw &  \targ & \targ & \targ & \targ & \gate{M_z} & \qw& \qw& \qw\\
	 }
 \]
\vspace{2mm}
\[
  \hspace{14mm}
    \Qcircuit @C=.92em @R=0.3em @!R {
	 	\lstick{q_{0}: \ket{+1}} & \gate{u_2(0,\frac{\pi}{2})}  & \ctrl{2} & \qw & \ctrl{2} & \qw & \qw & \qw& \qw\\
	 	\lstick{q_{1}: \ket{+1}} & \gate{H} & \qw & \ctrl{1} & \qw & \ctrl{1} & \qw & \qw & \qw\\
	 	\lstick{q_{2}: \ket{+1}} & \qw & \targ & \targ & \targ & \targ & \gate{M_z} & \qw& \qw\\
	 }
 \]
\vspace{2mm}
\[
  \hspace{14mm}
    \Qcircuit @C=.3em @R=.3em @!R {
	 	\lstick{q_{0}: \ket{+1}} & \ctrl{2} & \gate{H} & \qw & \ctrl{2} & \qw & \gate{u_{2}(\frac{\pi}{2},\frac{3\pi}{2})} & \ctrl{2} & \qw & \qw & \qw  \\
	 	\lstick{q_{1}: \ket{+1}} & \qw & \ctrl{1} & \gate{H} & \qw  & \ctrl{1} & \gate{u_{2}(\frac{\pi}{2},\frac{3\pi}{2})} & \qw & \ctrl{1} & \qw & \qw\\
	 	\lstick{q_{2}: \ket{+1}} & \targ & \targ & \qw & \targ & \targ & \qw & \targ & \targ & \gate{M_z} & \qw\\
	 }
 \]
\caption{\label{Fig:QC} Quantum circuits in the $z$-basis representing, from top to bottom, row 1 to row 3 and column 1 to column 3 in the Mermin-Peres magic square, see table \ref{Table 1}.}
\end{figure}
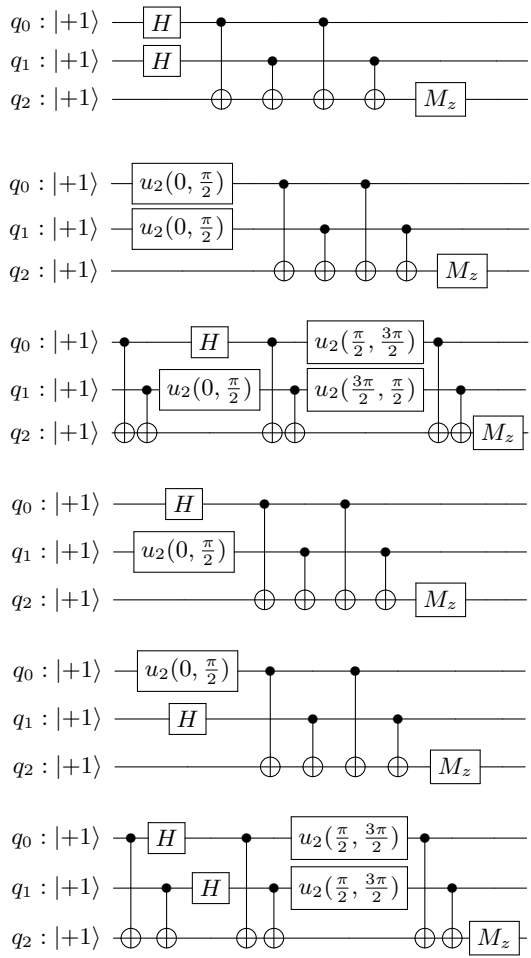

For each row, and for each column, a short program was written that included state preparation, coding the measurement results onto an ancillary qubit, and finally measuring the ancillary qubit to obtain the product of the measurement results. Since the qubits have significantly different performance, one needs to tailor the program so that the operations and the readout is made on the optimal qubit configuration. (We needed only three of the five qubits for our measurements.) It also helps to try to optimize the programs as much as possible, by carefully considering the operation order (since the three measurements done in every program commute) and by concatenating any series of gate operations that allows this. Running the same program on the same quantum computer, the results varied, for no apparent reason, more from day to day than one would predict from only statistical measurement fluctuations. The programs used to implement the six magic square measurement sequences are shown in standard quantum circuit notation in figure \ref{Fig:QC}. The programs and a short instruction are available online \cite{Dikme}.

Both the circuit concatenation and that only the product of the three measurement values is read out opens up for ``loopholes''. Ideally one would like to separate and read out the value of each measurement operator. Doing so one would also be able to switch the order of the measurements to certify that the measurements are indeed commutative. If not, so-called contextual biases may allow for a fully contextual model agreeing with the measured values. A detailed analysis of this situation can be found in \cite{Guhne,Kujala}. In our case, it is clear that switching the measurement sequence is unlikely to change anything for (top to bottom) circuit 1, 2, 4, and 5. However, for circuits 3 and 6 the six different, possible measurement sequences for each circuit will may modify the results. We tested if this was the case, and we found that indeed, we got worse results when e.g. row three of the magic square was measured in the order $\sx \otimes \sx$, $\sy \otimes \sy$, $\sz \otimes \sz$ rather than in the reverse order. The former measurement sequence require more gates (each generation errors) than the latter resulting in worse measurement statistics. On the other hand, the former measurement sequence resulted in almost the same measurement statistics as the sequence $\sx \otimes \sx$, $\sz \otimes \sz$, $\sy \otimes \sy$. This makes sense, since the number of needed gates is the same for these two measurement orders.

An even more problematic issue is the high readout errors. We did not read out the result of each individual observable in the square, but instead coded the measurement result onto an ancillary qubit and only read out the product of the three individual measurements by measuring this qubit. In theory, it would not matter if we read out the value of each separate measurement in the square and subsequently multiplied the measurement values, or if we simply coded the product of the values onto an ancillary qubit since the measurements in each row and in each column commute. In reality the multiple readout errors would significantly tip the measurement statistics towards an outcome explainable by realism and non-contextuality. Moreover, having only the product of the individual measurement outcomes makes it more difficult to check the commutativity (or more precisely, the contextual biases) of the measurements.

\section{Results}

As a pre-amble to measuring the full rows and columns of the magic square, we performed measurements of the nine individual observables on different input states on the ibmqx4 computer. For example, the results when measuring the input state $\ket{-1_z} \otimes \ket{-1_z}$ , where the notation  $-1_z$ means that it is the eigenstate to $\sz$ having eigenvalue $-1$, yielded the statistics in table \ref{Table 5} when repeated 8192 times (the highest number of repetitions IBM allows ordinary users).
\begin{table}
\caption{\label{Table 5} The number of +1 and -1 result outcomes for the nine measurements defined by the magic square depicted in table \ref{Table 1}. The frequency of the two possible outcomes for each entry in the magic square is given by the two numbers between a pair of double vertical lines. The input state was $\ket{-1_z} \otimes \ket{-1_z}$, and for example the frequencies 4440 and 3752 (top left) are the number of times (out of 8192) a measurement of $\sx \otimes 1$ yielded the result +1 and -1, respectively.}
\begin{tabular}{||c|c||c|c||c|c||}
  \hline
  +1 & -1 & +1 & -1 & +1 & -1 \\ \hline \hline
  4440 & 3752 & 4279 & 3913 & 4553 & 3639 \\ \hline
  4353 & 3839 & 4376 & 3816 & 4636 & 3556 \\ \hline
  4565 & 3627 & 4488 & 3704 & 7798 & 394 \\
  \hline
\end{tabular}
\end{table}
In theory, only the measurement of $\sz \otimes \sz$ (lower right) should produce a deterministic outcome, +1. The other eight measurements should yield the outcome +1 and -1 with equal probability. We see that this is not quite the case, the measurements have a slight bias toward the +1 outcome. Other input states such as $\ket{+1_z} \otimes \ket{+1_z}$, $\ket{+1_x} \otimes \ket{+1_y}$, and $\ket{+1_x} \otimes \ket{+1_x}$ produced very similar data, both for the measurements with nominally random outcomes and for that with a nominally deterministic outcome. The nominally random outcomes were always biased towards the +1 outcome. We did not systematically try with different entangled input states since we noted that significant errors were introduced already in the state preparation stage, and these errors could not be distinguished from the measurement errors.

Being stochastic measurements, we cannot expect the +1 and -1 results to occur exactly the same number of times even if they would have equal probability. The standard deviation of an equal probability, dichotomous random process repeated 8192 times is about 91, meaning that any number of outcomes for either result in the interval 4005 to 4187 would lie within plus or minus one standard deviation from the most likely result 8192/2=4096. We can note that the deviations in table \ref{Table 5} are larger than this, and they seem systematic. The nominally random outcome measurements were associated with a $\approx 8 \%$ bias towards +1, and had roughly the expected statistical variation around this bias. The nominally deterministic measurements had a $\approx 5 \%$ error. A plausible, partial explanation of this observation is that the state $\ket{+1}$ was internally represented on the quantum-computer transmon-qubit ground-state, whereas the state $\ket{-1}$ was coded onto an excited transmon state. Through dissipation, the latter state will relax to the former state, while the former suffers no dissipation, only dephasing, giving the qubits, and therefore the measurements, an asymmetry between the $\ket{+1}$ and $\ket{-1}$ states.

Having checked that individual measurements gave somewhat satisfactory results, we performed the six magic square joint measurements. Our best results were obtained on May 6, 2019 on the ibmqx4 computer. In a sequence of rerunning each program 8192 times, using the input state $\ket{+1_z}\otimes\ket{+1_z}$, we obtained the results in table \ref{Table 3}
\begin{table}
\caption{\label{Table 3} The measurement outcome of 8192 runs of each of the six programs implementing row 1 to column 3 of the Mermin-Peres magic square. ``Mean'' denotes the mean measured value, and ``Std dev'' denotes the standard deviation of this mean.}
\begin{tabular}{|l|c|c|c|c|}
  \hline
   & \multicolumn{2}{l|}{\# of outcomes}& Mean & Std dev  \\
   & 1 & -1 &   & $\times 10^{-3}$ \\ \hline
  Row 1 & 7943 & 249 & 0.939 & 3.79  \\ \hline
  Row 2 & 7731 & 461 & 0.887 & 5.09 \\ \hline
  Row 3 & 7506 & 686 & 0.833 & 6.12 \\ \hline
  Col 1 & 7813 & 379 & 0.907 & 4.64 \\ \hline
  Col 2 & 7851 & 341 & 0.917 & 4.41 \\ \hline
  Col 3 & 2033 & 6159 & -0.504 & 9.55 \\
  \hline
\end{tabular}
\end{table}

\section{Analysis}
As mentioned above, if one assumes that the magic square results can be explained by a realistic and non-contextual theory, then any result vector must be expressible as a statistical mixture of the 32 vectors $\overline{r}_j$, $j=1,2, \ldots, 32$. It follows that the scalar product between any result vector $\overline{v}$ satisfying realism and non-contextuality and any of the 32 vectors $\overline{q}_j$ satisfies
\begin{equation}
\overline{v} \cdot \overline{q}_j \leq 4 \quad \quad \forall j=1, 2, \ldots, 32.
\end{equation}
The proof of this assertion is simple. Since any scalar product $\overline{r}_j \cdot \overline{q}_k \leq 4$ $\forall j,k = 1,2, \ldots, 32$, this must also hold for any statistical mixture of the $\overline{r}_j$ vectors. On the other hand, for all the 32 vectors $\overline{q}_j$ it holds that $\overline{q}_j \cdot \overline{q}_j =6$, so quantum mechanics, predicting the result vector $\overline{q}_1$ for the magic square, allows results that violate the assumption of realism and non-contextuality.

The best result vector we obtained with the ibmqx4 machine gave the result vector $\overline{v}_b=(0.939,0.887,0.883,0.907,0.917, -0.504)$, when averaged over 8192 runs of each program representing a row of column measurement, see table \ref{Table 3}. Hence we find that $\overline{v} \cdot \overline{q}_1 = 4.987 \geq 4$. Thus, our results cannot be explained by a realistic and non-contextual theory.

To quantify this result, we estimate the standard deviation of each each of the averaged values. Since we deal with a dichotomous outcome distribution for each coefficient in the result vector the standard deviation of one run of a program can be estimated as
\begin{equation}
\delta=2\sqrt{p_+ (1-p_+)},
\end{equation}
where $p_+ $ is the estimated probability of getting the product of the measurement result in the row/column to +1. For $n$ runs of the program, the standard deviation of the average result thus will be
\begin{equation}
\delta=2\sqrt{\frac{p_+ (1-p_+)}{n}},
\end{equation}
assuming that each run yields a statistically independent error. (The systematic bias is incorporated in the average result.) The standard deviations are listed in the fifth column of table \ref{Table 3}. For later convenience we write them as an vector associated to $\overline{v}_b$ as $\overline{\delta}=(3.79,5.09,6.12,4.64,4.41,9.55)\times 10^{-3}$.

The possible outcome vectors for a theory that assumes realism and non-contextuality lies in or on the surface of the convex hull spanned by the vectors $\overline{r}_j$. Using the SciPy program library we used Delaunay triangulation to obtain the convex hull and constrained nonlinear optimization to find the closest point (defined by the Euclidean distance) on the hull to the result vector $\overline{v}_b$. This point is $(0.7466,0.7226,0.6685,0.7426,0.7526,-0.3390)$ and the Euclidean distance between $\overline{v}_b$ and the surface of the hull is 0.4029.

\begin{figure}
\includegraphics[width=0.6\columnwidth]{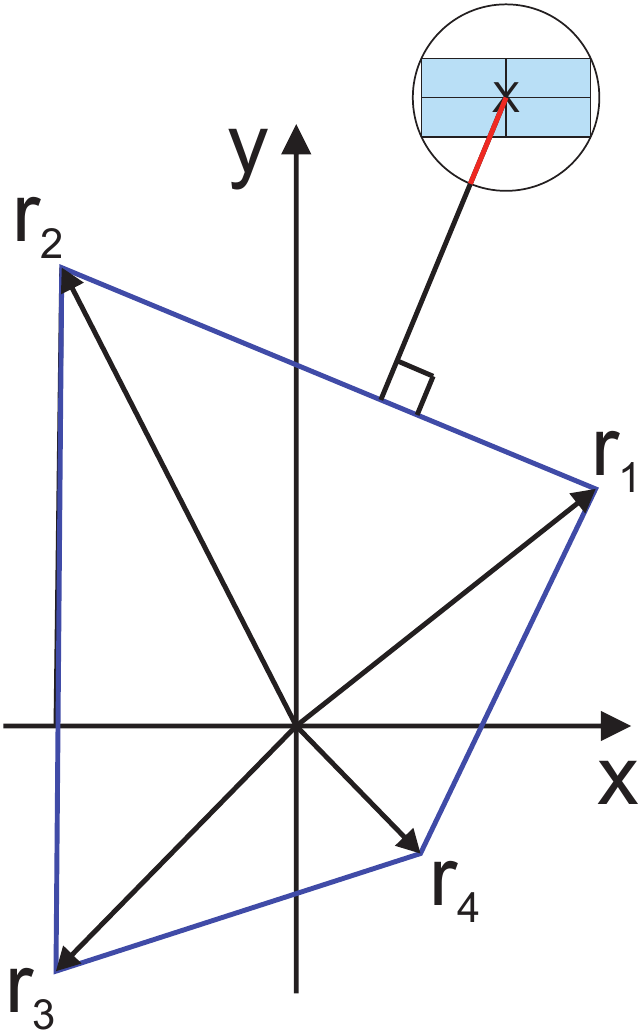}
\caption{\label{Fig:Standard_deviation} A schematic of a convex hull in the $xy$-plane, spanned by the realism result vectors $r_1$ to $r_4$. The X denotes the measurement outcome point (the tip of the result vector), and the width (height) of the light blue rectangle defines one standard deviation of the measurement in the $x$ ($y$) direction. The rectangle is inscribed in a standard deviation hypersphere (a circle in 2D). Its radius, red line, is a worst case estimate of the measurement point being the result of one standard deviation error in each direction in the plane. The black and red line gives the length from the measurement outcome point to the closest point on the convex hull.}
\end{figure}

To be conservative, one can form a ``standard deviation sphere'', a six-dimensional hypersphere with its radius defined by
\begin{equation}
r_\delta = \sqrt{\overline{\delta} \cdot \overline{\delta}}=0.0145.
\end{equation}
We find that our measured result vector violates the realistic and non-contextual hypothesis with at least $0.4029/0.0145 = 27.8$ standard deviations, see figure \ref {Fig:Standard_deviation} for a schematic illustration.

\section{Conclusions}
We have used an online, publicly available quantum computer to make the measurements defined in the Mermin-Peres magic square. Our result very convincingly supports the predictions of quantum mechanics and violate the bounds imposed by realistic and non-contextual theories. Of course, such a conclusion rests on certain assumptions. It has, e.g., been shown that the magic square can be simulated by a classical, three-state model, but for this model to be compatible with quantum predictions the model needs to keep track of what state will give correct predictions for a yet unperformed measurement given its past measurement and outcome \cite{Kleinmann}. Thus such a model necessitates a memory holding at least $\log_2(3)$ bits of information. In our experiment we have assumed that no such (classical) memory exist but that the ``memory'' of the past is carried solely by the measured qubits themselves.

Contemporary quantum computers are often characterized as NISQ (Noisy Intermediate-Scale Quantum) computers. Our study supports this nomenclature, because not only are the quantum computers we have used very small, they are also rather ``noisy'' in that unexpected results, due to decoherence of the qubits, imperfect gate fidelities, and qubit readout errors, are quite common. In about 10-30 \% of our program runs (depending on the program and choice of quantum computer qubits) the computers returned other results than those quantum mechanics predicts assuming ideal measurements. We also experienced that there were day-to-day fluctuations in the obtained results, and that the computers used delivered results of different quality. Occasionally the online quantum computers are ``tuned up'' and presumably the results are better if a program is run soon after such a ``tune-up''.

The imperfections notwithstanding, using quantum computers for quantum measurements holds a great promise. The qubits in contemporary quantum computers are getting better and better in that both the decoherence times and the gate fidelities are improving, and the readout errors are decreasing. For anyone who, like us, is using quantum computers for measurements, one wishes that especially the readout errors would decrease. For short programs, these errors are rather significant compared to the other errors. None-the-less, our work shows that an experiment that ten years ago could only be carried out in a highly specialized laboratory at a high cost is now at the fingertips of anyone in the world connected to the internet.

\section*{Acknowledgements}
This work was carried out with the support of the Swedish Research Council under grant 2014-7869-110765-32 and the Wallenberg Center for Quantum Technology. We acknowledge use of the IBM Quantum Experience. The views expressed are those of the authors and do not reflect the official policy or position of IBM or the IBM Quantum Experience team.

\end{document}